\documentclass{elsart}

\usepackage{amssymb}
\usepackage{graphicx}
\usepackage[ulem=normalem]{changes}

\begin{document}

\begin{frontmatter}



\title{Higher-rank discrete symmetries in the IBM.\\
I Octahedral shapes: general Hamiltonian}


\author[ganil]{P.~Van~Isacker,}
\author[batna]{A.~Bouldjedri}
and \author[batna]{S.~Zerguine}

\address[ganil]{Grand Acc\'el\'erateur National d'Ions Lourds, CEA/DSM--CNRS/IN2P3\\
Bd Henri Becquerel, BP 55027, F-14076 Caen Cedex 5, France}

\address[batna]{Department of Physics, PRIMALAB Laboratory, University of Batna\\
Avenue Boukhelouf M El Hadi, 05000 Batna, Algeria}

\begin{abstract}
In the context of the interacting boson model with $s$, $d$ and $g$ bosons,
the conditions for obtaining an intrinsic shape with octahedral symmetry
are derived for a general Hamiltonian with up to two-body interactions.
\end{abstract}


\end{frontmatter}

\section{Introduction}
\label{s_intro}
The collective model  of the atomic nucleus
assumes a description in terms of the shape of a surface
and its oscillations around that shape.
In nuclei quadrupole deformations with multipolarity $\lambda=2$
parameterize the most important deviation from spherical shape.
Their manifestation,
either as a vibration around a spherical shape
or through the mechanism of spontaneous symmetry breaking as a permanent deformation,
is by now a widely accepted feature of nuclear structure~\cite{BM75}.
Superimposed on quadrupole deformed shapes,
octupole ($\lambda=3$) and hexadecapole ($\lambda=4$) deformations
are frequently considered as well,
usually in terms of vibrational oscillations
although there is evidence for nuclei with a permanent octupole deformation~\cite{Gaffney13}.

The accepted paradigm is that
the spherical symmetry of the nucleus is reduced to a lower one
which in most cases corresponds to a shape with a single axis of symmetry.
In technical terms the spherical SO(3) symmetry
is spontaneously broken and reduced to SO(2).
Both SO(3) and SO(2) are examples of symmetries
that depend on {\em continuous} variables,
{\it e.g.}, the three Euler angles
or the angle of rotation around the axis of symmetry.
Nothing prohibits the further reduction of the symmetry
of the intrinsic shape of the nucleus
until a {\em discrete} (or {\em point}) symmetry
with a finite number of invariance operations remains
or even until no symmetry remains.
A well-known possibility is quadrupole deformation with three axes of different length.
The intrinsic nuclear shape in that case has no continuous symmetry
but is invariant under the eight transformations
that form the discrete group ${\rm D}_{\rm 2h}$.
(For a discussion of discrete groups see, {\it e.g.}, Hamermesh~\cite{Hamermesh64}.)
The invariance of the nuclear shape under rotations and reflections---or
the absence thereof in the case of deformation---determines the energy spectrum
and leads to predictions that can be tested experimentally.

In 1994 Li and Dudek~\cite{Li94} pointed out
that intrinsic shapes with a higher-rank discrete symmetry
can be obtained in the context of the collective model
and that, specifically, an octupole deformation with non-zero $\mu=2$ component
(and all other multipoles zero)
exhibits the tetrahedral symmetry ${\rm T}_{\rm d}$.
In subsequent studies~\cite{Dudek02,Dudek03,Dudek06,Rouvel14}
this observation was followed up systematically
and it was shown how the tetrahedral, octahedral and icosahedral discrete symmetries,
${\rm T}_{\rm d}$, ${\rm O}_{\rm h}$ and ${\rm I}_{\rm h}$,
arise through combinations of deformations of specific multipolarity.

For a fermionic quantum system,
{\it e.g.}, an odd-mass nucleus or a single-particle nuclear Hamiltonian,
time-reversal invariance must be considered as well.
Due to Kramers' theorem, stationary eigenstates of such systems
are at least twofold degenerate~\cite{BM69}.
Inclusion of the operation of time reversal
into the higher-rank discrete symmetries considered above
leads to the enlarged versions
${\rm T}_{\rm d}^{\rm D}$, ${\rm O}_{\rm h}^{\rm D}$ and ${\rm I}_{\rm h}^{\rm D}$
(see \S~99 in Landau and Lifchitz~\cite{Landau67}
who denote these enlarged discrete groups with a prime).
A prominent consequence of discrete symmetries of higher rank
is the occurrence of more than twofold degenerate states
in the spectrum of a single-particle nuclear Hamiltonian~\cite{Li94}.

While the formal possibility of nuclear shapes
with higher-rank discrete symmetries is by now well established,
the question remains whether such exotic deformations are realized in nuclei.
Over the years this question has been studied
from a theoretical~\cite{Takami98,Yamagami01,Schunck04}
and an experimental~\cite{Curien10,Curien11} point of view.
The former studies have been consistently carried out
in a mean-field approach usually supplemented with pairing correlations.
Concerning the experimental studies,
it is fair to say that up to now
no conclusive evidence has been found
that unambiguously establishes
the existence of a nucleus with a higher-rank discrete symmetry.
A notable example is the study of Jentschel {\it et al.}~\cite{Jentschel10}
who failed to find a vanishing quadrupole moment
of a negative-parity band in $^{156}$Gd
which should have been the `smoking gun' of tetrahedral deformation.

On the other hand, discrete symmetries are rather well established in light nuclei
in connection with alpha-particle clustering
which itself has a long history in nuclear physics~\cite{Brink66}.
Algebraic models have been developed by Bijker and Iachello~\cite{Bijker00,Bijker02}
with the aim to describe the discrete symmetries ${\rm D}_{\rm 3h}$ and ${\rm T}_{\rm d}$
associated with alpha-particles in a triangular or tetrahedral configuration.
This approach has attracted renewed interest
in view of novel experimental evidence
recently found in $^{12}$C and $^{16}$O~\cite{Bijker14,Marin14}.
Discrete symmetries associated with alpha-particle clustering,
while an important and attractive field of activity,
differ from those considered here
where they arise in the context of the collective model of the nucleus.

The aim of this series of papers is to analyze
the question of the possible occurrence of higher-rank discrete symmetries in nuclei
from a different theoretical perspective.
Since the work of Arima and Iachello~\cite{Arima75}
it is known that an alternative description of collective states in nuclei
exists in terms of bosons in the context of the interacting boson model (IBM).
Quadrupole collective states require $s$ and $d$ bosons,
with angular momentum $\ell=0$ and $\ell=2$, respectively,
and lead to the most elementary version of the model, the \mbox{$sd$-IBM}.
Many refinements of this original version are possible~\cite{Iachello87}
and already in the early papers on the IBM an $f$ boson ($\ell=3$)
is added to deal with negative-parity states with octupole collectivity~\cite{Arima76,Arima78,Scholten78}.
Hexadecapole states, on the other hand, require the consideration of a $g$ boson with $\ell=4$.

It will be shown that the higher-rank discrete symmetries,
as encountered in mean-field approaches,
can also be realized in the context of an algebraic model with the relevant degrees of freedom
and that, for example, tetrahedral and octahedral symmetries can be studied
in the context of the \mbox{$sf$-IBM} and \mbox{$sg$-IBM}, respectively.
Due to the pervasiveness of quadrupole collectivity in nuclei,
the addition of a $d$ boson will make these algebraic models less schematic,
leading to the \mbox{$sdf$-IBM} and \mbox{$sdg$-IBM}, respectively.
(In the former case it might even be necessary to consider the \mbox{$spdf$-IBM}
with an additional negative-parity $p$ boson with $\ell=1$~\cite{Engel85}.)

This series starts with an investigation of octahedral symmetries
in the framework of the \mbox{$sdg$-IBM},
adopting the model's most general Hamiltonian
with up to two-body interactions.
The collective parameters of quadrupole and hexadecapole deformation
and their relation to octahedral symmetry
are recalled in Sect.~\ref{s_shapes}.
In Sect.~\ref{s_sdgibm} the general Hamiltonian of the \mbox{$sdg$-IBM} is defined
and its corresponding classical limit in the most general coherent state is derived.
The catastrophe analysis of the resulting energy surface
is carried out in Sect.~\ref{s_octa}
with particular attention to the occurrence of minima with octahedral symmetry.
First conclusions from this analysis are drawn in Sect.~\ref{s_conc}.

\section{Quadrupole and hexadecapole shapes}
\label{s_shapes}
Shapes with octahedral discrete symmetry occur in lowest order
through a combination of hexadecapole deformations $Y_{4\mu}(\theta,\phi)$
with different $\mu$.
To make the shape more realistic for nuclei,
a quadrupole deformation should be added
since that deformation is of lowest order in the geometric model of Bohr.
With both these deformations the nuclear surface
is parameterized in the following way: 
\begin{eqnarray}
R(\theta,\phi)&=&
R_0\Bigl(1+
a_{20}Y_{20}(\theta,\phi)+
a_{22}[Y_{2-2}(\theta,\phi)+Y_{2+2}(\theta,\phi)]
\nonumber\\&&\phantom{R_0\Bigl[1}+
a_{40}Y_{40}(\theta,\phi)+
a_{42}[Y_{4-2}(\theta,\phi)+Y_{4+2}(\theta,\phi)]
\nonumber\\&&\phantom{R_0\Bigl[1+a_{40}Y_{40}(\theta,\phi)}+
a_{44}[Y_{4-4}(\theta,\phi)+Y_{4+4}(\theta,\phi)]\Bigr).
\label{e_surface1}
\end{eqnarray}
It is customary to define quadrupole-deformation variables
through
\begin{equation}
a_{20}=\beta_2\cos\gamma_2,
\qquad
a_{22}=\sqrt{\textstyle{\frac 1 2}}\beta_2\sin\gamma_2,
\label{e_param0}
\end{equation} 
where $\beta_2$ quantifies the quadrupole deviation from a sphere
and $\gamma_2$ the deviation from a quadrupole shape with axial symmetry.
Similarly, a hexadecapole variable $\beta_4$ is introduced
which quantifies the deviation from a sphere.
For the parameterization of hexadecapole asymmetric shapes
two approaches have been adopted.
In the first, a {\em single} $\gamma$ is introduced
which parameterizes the deviation from axial symmetry
at once for the quadrupole and hexadecapole degrees of freedom~\cite{Nazarewicz81}.
A symmetry argument then leads to a simplified parameterization
in terms of three variables $\beta_2$, $\beta_4$ and $\gamma$,
\begin{equation}
a_{40}=\sqrt{\textstyle{\frac 1 6}}\beta_4(5\cos^2\gamma+1),
\quad
a_{42}=\sqrt{\textstyle{\frac{15}{72}}}\beta_4\sin 2\gamma,
\quad
a_{44}=\sqrt{\textstyle{\frac{35}{72}}}\beta_4\sin^2\gamma.
\label{e_param1}
\end{equation}
In fact, there are three different such parameterizations~\cite{Nazarewicz81}
but Eq.~(\ref{e_param1}) is the one that has been used up to now
in the analysis of the \mbox{$sdg$-IBM} (see below).
The parameter ranges are
$0\leq\beta_2<+\infty$,
$-\infty<\beta_4<+\infty$
and $0\leq\gamma\leq\pi/3$,
and the values $\gamma=0$ and $\gamma=\pi/3$
lead to a shape with axial symmetry.

In a second approach,
developed by Rohozi\'nski and Sobiczewski~\cite{Rohozinski81},
a separate, independent asymmetry parameter $\gamma_4$ 
is introduced for the hexadecapole deformation.
Furthermore, to describe the full range of possible hexadecapole deformations,
an additional variable $\delta_4$ is needed,
which represents the convexity or concavity of the shape.
This leads to five shape variables,
two quadrupole and three hexadecapole ones
which are related as follows to the original parameterization~(\ref{e_surface1}):
\begin{eqnarray}
a_{40}&=&\beta_4\left(
\sqrt{\textstyle{\frac{7}{12}}}\cos\delta_4+
\sqrt{\textstyle{\frac{5}{12}}}\sin\delta_4\cos\gamma_4\right),
\nonumber\\
a_{42}&=&-\sqrt{\textstyle{\frac 1 2}}\beta_4\sin{\delta_4\sin\gamma_4},
\nonumber\\
a_{44}&=&\beta_4\left(
\sqrt{\textstyle{\frac{5}{24}}}\cos\delta_4-
\sqrt{\textstyle{\frac{7}{24}}}\sin\delta_4\cos\gamma_4\right),
\label{e_param2}
\end{eqnarray}
where the parameter ranges
are now $0\leq\beta_4<+\infty$,
$0\leq\gamma_4\leq\pi/3$
and $0\leq\delta_4\leq\pi$.
An axially symmetric shape occurs for $a_{42}=a_{44}=0$.
This corresponds to $\gamma_4=0$ and $\delta_4=\arccos\sqrt{7/12}$.

\begin{figure}
\centering
\includegraphics[width=5cm]{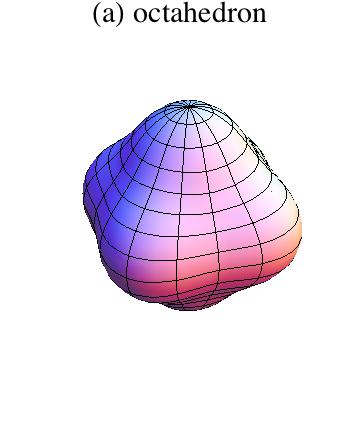}
\includegraphics[width=5cm]{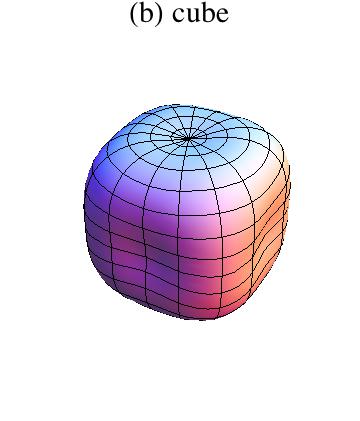}\\
\includegraphics[width=5cm]{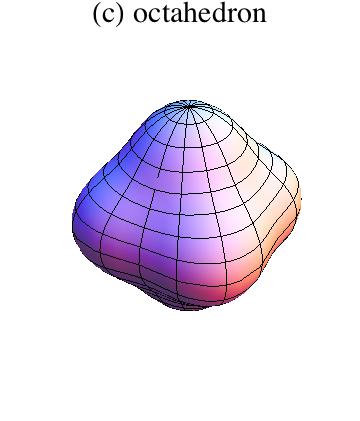}
\includegraphics[width=5cm]{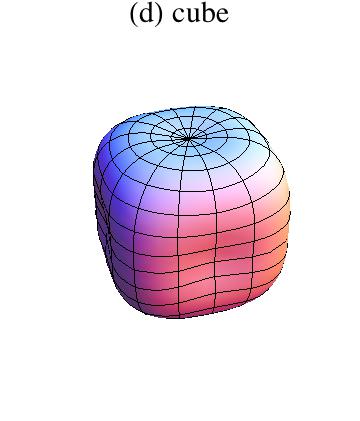}
\caption{Hexadecapole deformed surfaces with octahedral symmetry.
Four cases are shown:
(a) and (c) are octahedrons with $a_{40}=0.2$;
(b) and (d) are cubes with $a_{40}=-0.2$.
Octahedral symmetry is obtained
for $a_{42}=0$ and $a_{44}/a_{40}=+\sqrt{5/14}$ [(a) and (b)],
or $a_{42}=0$ and $a_{44}/a_{40}=-\sqrt{5/14}$ [(c) and (d)].
The two signs of $a_{44}/a_{40}$ correspond to the same intrinsic shape
but rotated over $\pi/2$ around the $z$ axis.
In terms of the parameters in Eq.~(\ref{e_param2}):
(a) $\beta_4=0.2$ and $\delta_4=0$,
(b) $\beta_4=0.2$ and $\delta_4=\pi$,
(c) $\beta_4=0.2$, $\delta_4=\arccos(1/6)$ and $\gamma_4=0$,
(d) not possible.}
\label{f_octa}
\end{figure}

A shape with octahedral symmetry implies a vanishing quadrupole deformation, $a_{20}=a_{22}=0$,
and can be realized in lowest order with a hexadecapole deformation
that satisfies~\cite{Dudek03,Dudek10}
\begin{equation}
a_{42}=0,
\qquad
a_{44}/a_{40}=\pm\sqrt{5/14}.
\label{e_condocta}
\end{equation}
The nuclear surface~(\ref{e_surface1}) then reduces to
\begin{equation}
R(\theta,\phi)=R_0
\left(1 + a_{40}\left[Y_{40}(\theta,\phi) \pm
\sqrt{\textstyle{\frac{5}{14}}}[Y_{4-4}(\theta,\phi)+Y_{4+4}(\theta,\phi)]\right]\right).
\label{e_surface2}
\end{equation}
Such shapes cannot be generated
with the restricted parameterization~(\ref{e_param1}),
which therefore is insufficient for the present purpose.
For positive values of $a_{40}$ an octahedron is obtained
while for negative $a_{40}$ one finds a cube, the dual of the octahedron,
both shapes having octahedral symmetry.
The sign of the ratio $a_{44}/a_{40}$ does not affect the intrinsic shape;
the opposite sign corresponds to the same shape
rotated over $\pi/2$ around the $z$ axis.
The four different cases are illustrated in Fig.~\ref{f_octa}.

In the parameterization~(\ref{e_param2})
the first of the octahedral conditions~(\ref{e_condocta})
implies $\delta_4=0$, $\delta_4=\pi$ or $\gamma_4=0$.
For $\delta_4=0$ or $\delta_4=\pi$,
the second of the octahedral conditions~(\ref{e_condocta})
is automatically satisfied for any value of $\gamma_4$,
resulting in a positive ratio $a_{44}/a_{40}$;
if $\delta_4=0$, the shape is an octahedron
while for $\delta_4=\pi$ it is a cube.
For the remaining case of $\gamma_4=0$
one finds an additional solution with octahedral symmetry for $\delta_4=\arccos(1/6)$,
corresponding to a rotated octahedron, see Fig.~\ref{f_octa}c,
but no rotated cube can be obtained with the parameterization~(\ref{e_param2}).

In the quadrupole parameterization~(\ref{e_param0})
a given couple $(\beta_2,\gamma_2)$
with $\beta_2\in[0,+\infty[$ and $\gamma_2\in[0,\pi/3]$
corresponds to a unique intrinsic shape.
This is not the case for the hexadecapole parameterization~(\ref{e_param2}),
that is, different triplets $(\beta_4,\delta_4,\gamma_4)$
with $\beta_4\in[0,+\infty[$, $\delta_4\in[0,\pi]$  and $\gamma_4\in[0,\pi/3]$
may lead to the same intrinsic shape,
differently oriented with respect to the laboratory frame.
An example of the latter are the triplets
$(\beta_4,\delta_4=0,\gamma_4={\rm anything})$
and $(\beta_4,\delta_4=\arccos(1/6),\gamma_4=0)$
which correspond to the same intrinsic shape for any value of $\beta_4$.
Shapes that are intrinsically the same
lead to identical conditions on the interaction parameters in the Hamiltonian,
as will be shown below.

\section{The $sdg$ interacting boson model}
\label{s_sdgibm}
Since the bosons of the IBM represent pairs of valence nucleons,
a nucleus is characterized by a constant total number of bosons $N$
which equals half the number of valence nucleons.
An important feature of the \mbox{$sd$-IBM} is the existence of a U(6) dynamical algebra,
the substructure of which leads to analytically solvable limits,
also called dynamical symmetries. 
The \mbox{$sd$-IBM} can successfully
describe quadrupole collective states in even-even nuclei
but other features require an extension of the \mbox{$sd$-IBM}.
In particular, the hexadecapole degree of freedom
requires the introduction of a $g$ boson ($\ell=4$)
and the upgrade of the dynamical algebra from U(6) to U(15).
We do not cite here the many papers related to the \mbox{$sdg$-IBM}
but refer the reader to the excellent review by Devi and Kota~\cite{Devi92}.

\subsection{Hamiltonian of the $sdg$-IBM}
\label{ss_sdgham}
Since the Hamiltonian of the \mbox{$sdg$-IBM} conserves the total number of bosons,
it can be written in terms of the 225 operators $b_{\ell m}^\dag b_{\ell' m'}$
where $b_{\ell m}^\dag$ ($b_{\ell m}$) creates (annihilates)
a boson with angular momentum $\ell$ and $z$ projection $m$.
This set of 225 operators generates the Lie algebra U(15).
A boson-number-conserving Hamiltonian with up to two-body interactions is of the form
\begin{equation}
\hat H=\hat H_1+\hat H_2.
\label{e_ham}
\end{equation}
The first term is the one-body part
\begin{eqnarray}
\hat H_1&=&
\epsilon_s[s^\dag\times\tilde s]^{(0)}+
\epsilon_d\sqrt{5}[d^\dag\times\tilde d]^{(0)}+
\epsilon_g\sqrt{9}[g^\dag\times\tilde g]^{(0)}
\nonumber\\&\equiv&
\epsilon_s\,s^\dag\cdot\tilde s+
\epsilon_d\,d^\dag\cdot\tilde d+
\epsilon_g\,g^\dag\cdot\tilde g
\nonumber\\&\equiv&
\epsilon_s\hat n_s+
\epsilon_d\hat n_d+
\epsilon_g\hat n_g,
\label{e_ham1}
\end{eqnarray}
where the multiplication $\times$ refers to coupling in angular momentum
(shown as an upperscript in round brackets),
the dot $\cdot$ indicates a scalar product
and $\tilde b_{\ell m}\equiv(-)^{\ell-m}b_{\ell,-m}$.
Furthermore, $\hat n_\ell$ is the number operator for the $\ell$ boson
and the coefficient $\epsilon_\ell$ is its energy.
The second term in the Hamiltonian~(\ref{e_ham})
represents the two-body interaction
\begin{equation}
\hat H_2=
\sum_{\ell_1\leq\ell_2,\ell'_1\leq\ell'_2,L}
\frac{(-)^Lv^L_{\ell_1\ell_2\ell'_1\ell'_2}}{\sqrt{(1+\delta_{\ell_1\ell_2})(1+\delta_{\ell'_1\ell'_2})}}
[b^\dag_{\ell_1}\times b^\dag_{\ell_2}]^{(L)}\cdot
[\tilde b_{\ell'_2}\times\tilde b_{\ell'_1}]^{(L)},
\label{e_ham2}
\end{equation}
where the coefficients $v^L_{\ell_1\ell_2\ell'_1\ell'_2}$ are the interaction matrix elements
between normalized two-boson states, 
\begin{equation}
v^L_{\ell_1\ell_2\ell'_1\ell'_2}=
\langle\ell_1\ell_2;LM_L|\hat H_2|\ell'_1\ell'_2;LM_L\rangle.
\label{e_int}
\end{equation}
It is henceforth assumed that $\ell_1\leq\ell_2$ and $\ell'_1\leq\ell'_2$.

Once the single-boson energies $\epsilon_\ell$
and interaction matrix elements $v^L_{\ell_1\ell_2\ell'_1\ell'_2}$ are known,
the most general two-body $sdg$-Hamiltonian is uniquely determined.

\subsection{Classical limit of the $sdg$-IBM}
\label{ss_sdgclas}
The classical limit of any boson Hamiltonian
is defined as its expectation value
in a coherent state~\cite{Gilmore79}.
This yields a function of the deformation variables
which can be interpreted as a total energy surface
depending on these variables.
The method was first proposed
for the \mbox{$sd$-IBM}~\cite{Ginocchio80,Dieperink80}.
The extension to the \mbox{$sdg$-IBM}
was carried out by Devi and Kota~\cite{Devi90}
for the simplified parameterization~(\ref{e_param1}).

The generic form of the coherent state for the \mbox{$sdg$-IBM} is
\begin{equation}
|N;a_{2\mu},a_{4\mu}\rangle\propto
\Gamma(a_{2\mu},a_{4\mu})^N|{\rm o}\rangle,
\label{e_coh1a}
\end{equation}
where
\begin{equation}
\Gamma(a_{2\mu},a_{4\mu})=
s^\dag+
\sum_{\mu=0,2}a_{2\mu}(d^\dag_{-\mu}+d^\dag_{+\mu})+
\sum_{\mu=0,2,4}a_{4\mu}(g^\dag_{-\mu}+g^\dag_{+\mu}),
\label{e_coh1b}
\end{equation}
and $|{\rm o}\rangle$ is the boson vacuum.
The $a_{\lambda\mu}$ have the interpretation of shape variables
appearing in the expansion~(\ref{e_surface1}).
Since the deformation in the IBM is generated by the valence nucleons only,
in contrast to the geometric model of Bohr and Mottelson~\cite{BM75}
where it is associated with the entire nucleus,
the shape variables in both models are proportional but not identical~\cite{Ginocchio80b}.
In terms of the variables
in the parameterizations~(\ref{e_param0}), (\ref{e_param1}) and (\ref{e_param2}),
$\beta_2$ and $\beta_4$ are proportional in both models
while the angles $\gamma$ or $\gamma_2$, $\gamma_4$ and $\delta_4$
have an identical interpretation.

The coherent state based on the parameterization~(\ref{e_param1})
\begin{equation}
|N;\beta_2,\beta_4,\gamma\rangle=
\sqrt{\frac{1}{N!(1+\beta_2^2+\beta_4^2)^N}}
\Gamma(\beta_2,\beta_4,\gamma)^N|{\rm o}\rangle,
\label{e_coh2a}
\end{equation}
with
\begin{eqnarray}
\Gamma(\beta_2,\beta_4,\gamma)&=&
s^\dag+
\beta_2\Bigl[\cos\gamma\,d^\dag_0+
\sqrt{\textstyle{\frac 1 2}}\sin\gamma\,(d^\dag_{-2}+d^\dag_{+2})\Bigr]
\nonumber\\&&+
\beta_4\Bigl[\textstyle{{\frac 1 6}}(5\cos^2\gamma+1)g^\dag_0+
\sqrt{\textstyle{\frac{15}{72}}}\sin2\gamma\,(g^\dag_{-2}+g^\dag_{+2})
\nonumber\\&&\phantom{+\beta_4\Bigl[}+
\sqrt{\textstyle{\frac{35}{72}}}\sin^2\gamma\,(g^\dag_{-4}+g^\dag_{+4})\Bigr],
\label{e_coh2b}
\end{eqnarray}
was used by Devi and Kota~\cite{Devi90}
to derive the geometry of the different limits of the \mbox{$sdg$-IBM}.
The expression for the classical limit of the general \mbox{$sdg$-Hamiltonian}~(\ref{e_ham})
with this coherent state was given in Ref.~\cite{Isacker10}.

If one is interested in octahedral shapes
and how they appear in the \mbox{$sdg$-IBM},
the general parameterization~(\ref{e_param2}) is needed,
and the appropriate coherent state is
\begin{equation}
|N;\beta_2,\beta_4,\gamma_2,\gamma_4,\delta_4\rangle=
\sqrt{\frac{1}{N!(1+\beta_2^2+\beta_4^2)^N}}
\Gamma(\beta_2,\beta_4,\gamma_2,\gamma_4,\delta_4)^N|{\rm o}\rangle,
\label{e_coh3a}
\end{equation}
with
\begin{eqnarray}
\Gamma(\beta_2,\beta_4,\gamma_2,\gamma_4,\delta_4)&=&
s^\dag+
\beta_2\Bigl[\cos\gamma_2d^\dag_0+
\sqrt{\textstyle{\frac 1 2}}\sin\gamma_2(d^\dag_{-2}+d^\dag_{+2})\Bigr]
\label{e_coh3b}\\&&+
\beta_4\Bigl[\Bigl(\sqrt{\textstyle{\frac{7}{12}}}\cos\delta_4+
\sqrt{\textstyle{\frac{5}{12}}}\sin\delta_4\cos\gamma_4\Bigr)g^\dag_0
\nonumber\\&&\phantom{+\beta_4\Bigl[}-
\sqrt{\textstyle{\frac 1 2}}\sin\delta_4\sin\gamma_4(g^\dag_{-2}+g^\dag_{+2})
\nonumber\\&&\phantom{+\beta_4\Bigl[}+
\Bigl(\sqrt{\textstyle{\frac{5}{24}}}\cos\delta_4-
\sqrt{\textstyle{\frac{7}{24}}}\sin\delta_4\cos\gamma_4\Bigr)
(g^\dag_{-4}+g^\dag_{+4})\Bigr].
\nonumber
\end{eqnarray}

The classical limit of any Hamiltonian of the \mbox{$sdg$-IBM} is,
for the general coherent state~(\ref{e_coh3a}), defined as
\begin{equation}
\langle\hat H\rangle\equiv
\langle N;\beta_2,\beta_4,\gamma_2,\gamma_4,\delta_4|\hat H
|N;\beta_2,\beta_4,\gamma_2,\gamma_4,\delta_4\rangle.
\label{e_climit0}
\end{equation}
Once the form of the coherent state has been determined,
the expectation value~(\ref{e_climit0}) can be obtained
with the method of differentiation~\cite{Isacker81}
which lends itself ideally to programming in a symbolic language.
The classical limit of the one-body part~(\ref{e_ham1}) is
\begin{equation}
\langle\hat H_1\rangle=
N\frac{\epsilon_s+\epsilon_d\beta_2^2+\epsilon_g\beta_4^2}{1+\beta_2^2+\beta_4^2},
\label{e_climit1}
\end{equation}
while that of its two-body part~(\ref{e_ham2})
can be written in the generic form
\begin{equation}
\langle\hat H_2\rangle=
\frac{N(N-1)}{(1+\beta_2^2+\beta_4^2)^2}
\sum_{kl}\beta_2^k\beta_4^l
\left[c_{kl}+\sum_{ij}c^{ij}_{kl}\cos(i\gamma_2+j\gamma_4)\phi_{kl}^{ij}(\delta_4)\right],
\label{e_climit2}
\end{equation}
where the coefficients $c_{kl}$ and $c^{ij}_{kl}$
can be expressed in terms of the interactions $v^L_{\ell_1\ell_2\ell'_1\ell'_2}$.
The expressions for the non-zero coefficients $c_{kl}$ are
\begin{eqnarray}&&
\textstyle
c_{00}={\frac 1 2}v_{ssss}^0,
\quad
c_{20}=\sqrt{\frac 1 5}v_{ss\cdot dd}^0+v_{sdsd}^2,
\quad
c_{02}={\frac 1 3}v_{ss\cdot gg}^0+v_{sgsg}^4,
\nonumber\\&&
\textstyle
c_{40}=\frac{1}{10}v_{dddd}^0+{\frac 1 7}v_{dddd}^2+\frac{9}{35}v_{dddd}^4,
\nonumber\\&&
\textstyle
c_{22}=
\frac{1}{3\sqrt{5}}v_{dd\cdot gg}^0
+\frac{7}{\sqrt{715}}v_{dd\cdot gg}^4
+\frac{1}{6}v_{dgdg}^2
+\frac{4}{11}v_{dgdg}^4
+\frac{1}{6}v_{dgdg}^5
+\frac{10}{33}v_{dgdg}^6,
\nonumber\\&&
\textstyle
c_{04}=
\frac{1}{18}v_{gggg}^0
+\frac{38}{693}v_{gggg}^2
+\frac{89}{1001}v_{gggg}^4
+\frac{62}{495}v_{gggg}^6
+\frac{1129}{6435}v_{gggg}^8,
\label{e_coef2}
\end{eqnarray}
while those for the non-zero coefficients $c^{ij}_{kl}$ are
\begin{eqnarray}&&
\textstyle
c^{30}_{30}=-\frac{2}{\sqrt{7}}v_{sd\cdot dd}^2,
\nonumber\\&&
\textstyle
c^{00}_{21}=\sqrt{\frac 2 3}v_{sd\cdot dg}^2+\sqrt{\frac 3 5}v_{sg\cdot dd}^4,
\quad
c^{21}_{21}=\sqrt{\frac{10}{21}}v_{sd\cdot dg}^2+\sqrt{\frac 3 7}v_{sg\cdot dd}^4,
\nonumber\\&&
\textstyle
c^{1-1}_{12}=-\frac{2}{3}\sqrt{\frac{10}{11}}v_{sd\cdot gg}^2-\frac{4}{\sqrt{11}}v_{sg\cdot dg}^4,
\quad
c^{12}_{12}=\frac{2}{3}\sqrt{\frac{2}{77}}v_{sd\cdot gg}^2+\frac{4}{\sqrt{385}}v_{sg\cdot dg}^4,
\nonumber\\&&
\textstyle
c^{00}_{03}=\frac{2}{\sqrt{429}}v_{sg\cdot gg}^4,
\quad
c^{03}_{03}=\sqrt{\frac{80}{3003}}v_{sg\cdot gg}^4,
\nonumber\\&&
\textstyle
c^{1-1}_{31}=-\frac{1}{7}\sqrt{\frac{10}{3}}v_{dd\cdot dg}^2-\frac{2}{7}\sqrt{\frac{15}{11}}v_{dd\cdot dg}^4,
\quad
c^{30}_{31}=-\sqrt{\frac{2}{21}}v_{dd\cdot dg}^2-\sqrt{\frac{12}{77}}v_{dd\cdot dg}^4,
\nonumber\\&&
\textstyle
c^{00}_{22}=
-\frac{3}{\sqrt{715}}v_{dd\cdot gg}^4
-\frac{1}{42}v_{dgdg}^2
+\frac{1}{15}v_{dgdg}^3
-\frac{27}{385}v_{dgdg}^4
+\frac{1}{30}v_{dgdg}^5
-\frac{1}{165}v_{dgdg}^6,
\nonumber\\&&
\textstyle
c^{2-2}_{22}=
-\frac{2}{21}\sqrt{\frac{2}{11}}v_{dd\cdot gg}^2
+\frac{4}{7}\sqrt{\frac{5}{143}}v_{dd\cdot gg}^4
-\frac{1}{15}v_{dgdg}^3
+\frac{1}{11}v_{dgdg}^4
-\frac{4}{165}v_{dgdg}^6,
\nonumber\\&&
\textstyle
c^{21}_{22}=
\frac{2}{3}\sqrt{\frac{10}{77}}v_{dd\cdot gg}^2
+\sqrt{\frac{1}{1001}}v_{dd\cdot gg}^4
+\frac{1}{6}\sqrt{\frac{5}{7}}v_{dgdg}^2
-\frac{8}{11\sqrt{35}}v_{dgdg}^4
-\frac{1}{6}\sqrt{\frac{7}{5}}v_{dgdg}^5
\nonumber\\&&\phantom{c^{21}_{22}=}
\textstyle
+\frac{\sqrt{35}}{33}v_{dgdg}^6,
\nonumber\\&&
\textstyle
c^{1-1}_{13}=
-\frac{4}{21}\sqrt{\frac{5}{33}}v_{dg\cdot gg}^2
-\frac{12}{77}\sqrt{\frac{3}{13}}v_{dg\cdot gg}^4
-\frac{2}{33}\sqrt{\frac{10}{3}}v_{dg\cdot gg}^6,
\nonumber\\&&
\textstyle
c^{12}_{13}=
-{\frac{4}{3\sqrt{231}}}v_{dg\cdot gg}^2
-\frac{12}{11}\sqrt{\frac{3}{455}}v_{dg\cdot gg}^4
-\frac{2}{33}\sqrt{\frac{14}{3}}v_{dg\cdot gg}^6,
\nonumber\\&&
\textstyle
c^{00}_{04}=
-\frac{2}{693}v_{gggg}^2
+\frac{4}{3003}v_{gggg}^4
+\frac{2}{495}v_{gggg}^6
-\frac{16}{6435}v_{gggg}^8,
\nonumber\\&&
\textstyle
c^{03}_{04}=
-\frac{2}{99}\sqrt{\frac{5}{7}}v_{gggg}^2
+\frac{4}{429}\sqrt{\frac{5}{7}}v_{gggg}^4
+\frac{2}{99}\sqrt{\frac{7}{5}}v_{gggg}^6
-\frac{16}{1287}\sqrt{\frac{7}{5}}v_{gggg}^8,
\label{e_coef3}
\end{eqnarray}
where the notation
$v^L_{\ell_1\ell_2\cdot\ell'_1\ell'_2}\equiv
v^L_{\ell_1\ell_2\ell'_1\ell'_2}+v^L_{\ell'_1\ell'_2\ell_1\ell_2}$
is introduced for $(\ell_1\ell_2)\neq(\ell'_1\ell'_2)$
since this combination consistently occurs due to the hermiticity of the Hamiltonian.
Furthermore, the $\phi^{ij}_{kl}(\delta_4)$ are functions defined as follows:
\begin{eqnarray}&&
\phi^{30}_{30}(\delta_4)=1,
\quad
\phi^{00}_{21}(\delta_4)=\cos\delta_4,
\quad
\phi^{21}_{21}(\delta_4)=\sin\delta_4,
\nonumber\\&&
\phi^{1-1}_{12}(\delta_4)=\sin2\delta_4,
\quad
\phi^{12}_{12}(\delta_4)=1-\cos2\delta_4,
\nonumber\\&&
\phi^{00}_{03}(\delta_4)=6\cos\delta_4+\cos3\delta_4,
\quad
\phi^{03}_{03}(\delta_4)=3\sin\delta_4-\sin3\delta_4,
\nonumber\\&&
\phi^{1-1}_{31}(\delta_4)=\sin\delta_4,
\quad
\phi^{30}_{31}(\delta_4)=\cos\delta_4,
\nonumber\\&&
\phi^{00}_{22}(\delta_4)=\phi^{2-2}_{22}(\delta_4)=1-\cos2\delta_4,
\quad
\phi^{21}_{22}(\delta_4)=\sin2\delta_4,
\nonumber\\&&
\phi^{1-1}_{13}(\delta_4)=\sin\delta_4+2\sin3\delta_4,
\quad
\phi^{12}_{13}(\delta_4)=\cos\delta_4-\cos3\delta_4,
\nonumber\\&&
\phi^{00}_{04}(\delta_4)=2\cos2\delta_4+17\cos4\delta_4,
\quad
\phi^{03}_{04}(\delta_4)=2\sin2\delta_4-\sin4\delta_4.
\label{e_funcs}
\end{eqnarray}
These functions can be written concisely as
\begin{equation}
\phi^{ij}_{kl}(\delta_4)=
\sum_{n=l,l-2,\dots}^{0\;{\rm or}\;1}c^{ij}_{kln}\varphi_j(n\delta_4),
\end{equation}
where $\varphi_j(\theta)$ is $\cos\theta$ ($\sin\theta$) for even (odd) $j$,
with the following interaction-independent constants:
\begin{eqnarray}&&
c^{30}_{300}=1,
\quad
c^{00}_{211}=1,
\quad
c^{21}_{211}=1,
\nonumber\\&&
c^{1-1}_{120}=0,
\quad
c^{1-1}_{122}=1,
\quad
c^{12}_{120}=1,
\quad
c^{12}_{122}=-1,
\nonumber\\&&
c^{00}_{031}=6,
\quad
c^{00}_{033}=1,
\quad
c^{03}_{031}=3,
\quad
c^{03}_{033}=-1,
\nonumber\\&&
c^{1-1}_{311}=1,
\quad
c^{30}_{311}=1,
\nonumber\\&&
c^{00}_{220}=c^{2-2}_{220}=1,
\quad
c^{00}_{222}=c^{2-2}_{222}=-1,
\quad
c^{21}_{220}=0,
\quad
c^{21}_{222}=1,
\nonumber\\&&
c^{1-1}_{131}=1,
\quad
c^{1-1}_{133}=2,
\quad
c^{12}_{131}=1,
\quad
c^{12}_{133}=-1,
\nonumber\\&&
c^{00}_{040}=0,
\quad
c^{00}_{042}=2,
\quad
c^{00}_{044}=17,
\quad
c^{03}_{040}=0,
\quad
c^{03}_{042}=2,
\quad
c^{03}_{044}=-1.
\label{e_consts}
\end{eqnarray}
The classical limit of the total Hamiltonian~(\ref{e_ham})
can therefore be written as
\begin{eqnarray}
\langle\hat H\rangle&\equiv&
E(\beta_2,\beta_4,\gamma_2,\gamma_4,\delta_4)
\nonumber\\&=&
\frac{N(N-1)}{(1+\beta_2^2+\beta_4^2)^2}
\sum_{kl}\beta_2^k\beta_4^l
\left[c'_{kl}+\sum_{ij}c^{ij}_{kl}\cos(i\gamma_2+j\gamma_4)\phi_{kl}^{ij}(\delta_4)\right],
\label{e_climit}
\end{eqnarray}
where $c'_{kl}$ are the modified coefficients
\begin{eqnarray}&&
c'_{00}=c_{00}+\epsilon'_s,
\quad
c'_{20}=c_{20}+\epsilon'_s+\epsilon'_d,
\quad
c'_{02}=c_{02}+\epsilon'_s+\epsilon'_g,
\nonumber\\&&
c'_{40}=c_{40}+\epsilon'_d,
\quad
c'_{22}=c_{22}+\epsilon'_d+\epsilon'_g,
\quad
c'_{04}=c_{04}+\epsilon'_g,
\label{e_coef2a}
\end{eqnarray}
in terms of the scaled boson energies $\epsilon'_\ell\equiv\epsilon_\ell/(N-1)$.
While the differentiation technique~\cite{Isacker81}
allows a secure derivation of the expectation value~(\ref{e_climit0}),
the particular representation~(\ref{e_climit})
in terms of functions $\beta_2^k\beta_4^l\cos(i\gamma_2+j\gamma_4)\phi_{kl}^{ij}(\delta_4)$
is not obtained automatically.
The correctness of the latter representation can be proven
by use of trigonometric conversion algorithms
which show it to be identical to the expression
found with the brute-force differentiation technique.

The quantum-mechanical Hamiltonian~(\ref{e_ham}), if it is hermitian,
depends on three single-boson energies $\epsilon_\ell$
and 32 two-body interactions $v^L_{\ell_1\ell_2\ell'_1\ell'_2}$.
In the classical limit with the most general coherent state~(\ref{e_coh3a}),
the number of independent parameters
in the energy surface $E(\beta_2,\beta_4,\gamma_2,\gamma_4,\delta_4)$ is reduced to 22
(six coefficients $c'_{kl}$ and 16 coefficients $c^{ij}_{kl}$).
For comparison, if the simpler coherent state~(\ref{e_coh2a}) is taken~\cite{Isacker10},
this number is further reduced to 15.

\section{Octahedral shapes in the $sdg$-IBM}
\label{s_octa}
\subsection{Principle of the method}
\label{ss_princip}
It is clear that a catastrophe analysis of the energy surface
$E(\beta_2,\beta_4,\gamma_2,\gamma_4,\delta_4)$
with its five order parameters and 22 control parameters
is beyond the scope of any reasonable analysis.
Fortunately, this is not needed if one is interested
in the realization of octahedral symmetry in the \mbox{$sdg$-IBM}.
For this purpose one just wants to know what are the conditions
on the interactions in the \mbox{$sdg$-Hamiltonian}
for the surface~(\ref{e_climit}) to have a minimum with octahedral symmetry.
As shown in Sect.~\ref{s_shapes},
a shape with such symmetry occurs for
(i) $\beta_2=0$, $\beta_4\neq0$, $\gamma_2={\rm anything}$, $\gamma_4={\rm anything}$ and $\delta_4=0$,
(ii) $\beta_2=0$, $\beta_4\neq0$, $\gamma_2={\rm anything}$, $\gamma_4={\rm anything}$ and $\delta_4=\pi$,
or (iii) $\beta_2=0$, $\beta_4\neq0$, $\gamma_2={\rm anything}$, $\gamma_4=0$ and $\delta_4=\arccos(1/6)$.

The conditions for the energy surface
$E(\beta_2,\beta_4,\gamma_2,\gamma_4,\delta_4)$
to have an extremum at $p^*$ are
\begin{equation}
\left.\frac{\partial E}{\partial\beta_2}\right|_{p^*}=
\left.\frac{\partial E}{\partial\beta_4}\right|_{p^*}=
\left.\frac{\partial E}{\partial\gamma_2}\right|_{p^*}=
\left.\frac{\partial E}{\partial\gamma_4}\right|_{p^*}=
\left.\frac{\partial E}{\partial\delta_4}\right|_{p^*}=0,
\label{e_extr}
\end{equation}
where $p^*\equiv(\beta^*_2,\beta^*_4,\gamma^*_2,\gamma^*_4,\delta^*_4)$
is a short-hand notation for an arbitrary critical point.
Furthermore, a critical point with octahedral symmetry
shall be denoted as $o^*$ which implies that $o^*$
is one of the three cases (i), (ii) or (iii) listed above.
While the Eqs.~(\ref{e_extr}) are necessary
for $E(\beta_2,\beta_4,\gamma_2,\gamma_4,\delta_4)$ to have an {\em extremum} at $p^*$,
the conditions for a {\em minimum} require in addition
that the eigenvalues of the stability matrix
[{\it i.e.}, the partial derivatives of $E(\beta_2,\beta_4,\gamma_2,\gamma_4,\delta_4)$ of second order]
are all non-negative.
All expressions in this section are obtained
starting from the generic expression~(\ref{e_climit}) for the energy surface
and its derivatives up to second order.

\subsection{Extrema with octahedral symmetry}
\label{ss_extrocta}
Let us now apply the above procedure to the case of octahedral symmetry
which requires the establishment of an extremum
of the energy surface $E(\beta_2,\beta_4,\gamma_2,\gamma_4,\delta_4)$ for $p^*=o^*$.

Consider first the case of an octahedral shape (i) or a cubic shape (ii),
cases that can be treated simultaneously.
Four of the five extremum conditions~(\ref{e_extr})
are identically satisfied for $p^*=o^*$
and do not lead to any constraints on the coefficients $c'_{kl}$ and $c^{ij}_{kl}$.
The derivative in $\beta_4$ leads to the equation
\begin{equation}
\beta_4^*\left[
-4c'_{00}+2c'_{02}\pm
21c^{00}_{03}\beta_4^*-
\left(2c'_{02}-4c'_{04}-76c^{00}_{04}\right)\beta_4^{*2}\mp
7c^{00}_{03}\beta_4^{*3}
\right]=0,
\label{e_extb4}
\end{equation}
where the upper (lower) sign applies to $\delta_4^*=0$ ($\delta_4^*=\pi$).
The spherical point $\beta_4^*=0$ is always an extremum of the energy surface.
Other extrema $\beta_4^*$ are found as solutions of a cubic equation
and therefore have cumbersome expressions.
In the search for octahedral minima
one is not interested in numerical values or analytic expressions for $\beta_4^*$
but one simply wants to know
whether a hexadecapole deformed minimum exists or not.
This question can be readily answered
if one assumes the coefficients of the odd powers of $\beta_4^*$
in Eq.~(\ref{e_extb4}) to be zero,
which happens for $c^{00}_{03}=0$.
In that case, the non-zero solutions of Eq.~(\ref{e_extb4}) are
\begin{equation}
\beta_4^*=\pm\sqrt{\frac{2c'_{00}-c'_{02}}{-c'_{02}+2c'_{04}+38c^{00}_{04}}},
\label{e_solb4}
\end{equation}
leading to the conclusion that a real, positive $\beta_4^*$ is found
if the combinations $2c'_{00}-c'_{02}$ and $-c'_{02}+2c'_{04}+38c^{00}_{04}$
have the same sign.
For non-zero values of $c^{00}_{03}$
[which is related to the $s$-$g$ mixing matrix element $v_{sg\cdot gg}^4$,
see Eq.~(\ref{e_coef3})]
the analysis is more complicated.
The cubic equation~(\ref{e_extb4}) has real coefficients
and therefore it always has at least one real solution $\beta_4^*$.
In addition, since the ratio $\mp c^{00}_{03}/(-4c'_{00}+2c'_{02})$ is negative
for one of the choices $\delta_4^*=0$ or $\delta_4^*=\pi$,
it follows that the solution $\beta_4^*$ is positive in that case.
We conclude that there exists always an extremum with octahedral symmetry
for any \mbox{$sdg$-Hamiltonian}
except in the pathological case of no  $s$-$g$ mixing, $v_{sg\cdot gg}^4=0$,
in which case the condition is that
the combinations $2c'_{00}-c'_{02}$ and $-c'_{02}+2c'_{04}+38c^{00}_{04}$
should have the same sign.

For the case (iii) with $\delta_4=\arccos(1/6)$,
the derivatives in $\gamma_2$ and $\gamma_4$ are identically zero
and do not lead to any condition.
The derivatives in $\beta_2$, $\beta_4$ and $\delta_4$ lead to the equations
\begin{eqnarray}&&
\beta_4^{*2}\left[
-3\left(\sqrt{35}c^{1-1}_{12}+35c^{12}_{12}\right)+
7\left(\sqrt{35}c^{1-1}_{13}-5c^{12}_{13}\right)\beta_4^*
\right]=0,
\nonumber\\&&
\beta_4^*\left[
324\left(-2c'_{00}+c'_{02}\right)+
63\left(4c^{00}_{03}+5\sqrt{35}c^{03}_{03}\right)\beta_4^*
\right.
\nonumber\\&&\quad+
4\left(-81c'_{02}+162c'_{04}+1853c^{00}_{04}+35\sqrt{35}c^{03}_{04}\right)\beta_4^{*2}
\nonumber\\&&\quad-\left.
21\left(4c^{00}_{03}+5\sqrt{35}c^{03}_{03}\right)\beta_4^{*3}
\right]=0,
\nonumber\\&&
\beta_4^{*3}\left[
9\left(-2\sqrt{35}c^{00}_{03}+7c^{03}_{03}\right)+
224\left(\sqrt{35}c^{00}_{04}-c^{03}_{04}\right)\beta_4^*
\right]=0.
\label{e_extb2b4d4}
\end{eqnarray}
It would therefore seem that for the energy surface~(\ref{e_climit})
the critical conditions in the case (iii) lead to equations
that are different from the those obtained in the cases (i) and (ii).
It should not be forgotten, however, that the coefficients $c'_{kl}$ and $c^{ij}_{kl}$
are expressed in terms of single-boson energies $\epsilon_\ell$
and interaction matrix elements $v^L_{\ell_1\ell_2\ell'_1\ell'_2}$.
After substitution of the expressions given in Eqs.~(\ref{e_coef2}), (\ref{e_coef3}) and~(\ref{e_coef2a}),
it is found that the first and third conditions of Eq.~(\ref{e_extb2b4d4}) are identically satisfied
while the second reduces to the one given in Eq.~(\ref{e_extb4}) with the upper sign.
This confirms the earlier statement that
a given intrinsic shape leads to unique conditions
on the single-boson energies and interaction matrix elements,
independent of the orientation of that shape in the laboratory frame
The result provides an additional and independent check on the correctness
of all the equations involved in this comparison.
Therefore, the analysis henceforth
can be restricted to the cases (i) and (ii) of octahedral and cubic intrinsic shape.
 
\subsection{Minima with octahedral symmetry}
\label{ss_minocta}
So far, no constraints are found
on the single-boson energies $\epsilon_\ell$
and interaction matrix elements $v^L_{\ell_1\ell_2\ell'_1\ell'_2}$
since Eq.~(\ref{e_extb4}) has always a real, positive solution,
either for $\delta_4^*=0$ or for $\delta_4^*=\pi$,
except in the pathological case of no  $s$-$g$ mixing
mentioned in Sect.~\ref{ss_extrocta}.
Constraints are found by requiring that the extremum is a minimum
or, equivalently, that the eigenvalues of the stability matrix are all non-negative.
No conditions follow from second derivatives involving $\gamma_2$ and $\gamma_4$
at a critical point $o^*$ with octahedral symmetry.
Furthermore, the second derivatives involving $\beta_4$
are decoupled from those pertaining to $\beta_2$ and $\delta_4$,
that is, the following equations are identically satisfied:
\begin{equation}
\left.\frac{\partial^2 E}{\partial\beta_2\partial\beta_4}\right|_{o^*}=
\left.\frac{\partial^2 E}{\partial\beta_4\partial\delta_4}\right|_{o^*}=0,
\label{e_stab1}
\end{equation}
so that the stability in $\beta_4$ follows from the inequality
\begin{equation}
\left.\frac{\partial^2 E}{\partial\beta_4^2}\right|_{o^*}\geq0.
\label{e_stab2}
\end{equation}
Some insight can be obtained
by assuming the odd powers of $\beta_4^*$ to be zero, $c^{00}_{03}=0$,
in which case the condition~(\ref{e_stab2}) reduces to
\begin{equation}
\frac{(2c'_{00}-c'_{02})(c'_{00}-c'_{02}+c'_{04}+19c^{00}_{04})}
{-c'_{02}+2c'_{04}+38c^{00}_{04}}\geq0.
\label{e_stab3}
\end{equation}
Since the combinations $2c'_{00}-c'_{02}$ and $-c'_{02}+2c'_{04}+38c^{00}_{04}$
must have the same sign (see Sect.~\ref{ss_extrocta}),
it follows that the conditions
\begin{equation}
2c'_{00}-c'_{02}\geq0,
\quad
-c'_{02}+2c'_{04}+38c^{00}_{04}\geq0.
\label{e_cond1}
\end{equation}
are necessary and sufficient for the energy surface $E(\beta_2,\beta_4,\gamma_2,\gamma_4,\delta_4)$
to have an extremum at non-zero hexadecapole deformation
which is stable in $\beta_4$.

The stability in $\beta_2$ and $\delta_4$ 
follows from the diagonal derivatives,
\begin{eqnarray}
\left.\frac{\partial^2 E}{\partial\beta_2^2}\right|_{o^*}&=&
\frac{-4c'_{00}+2c'_{20}\pm
2c^{00}_{21}\beta_4^*-
(4c'_{02}-2c'_{20}-2c'_{22})\beta_4^{*2}}
{\phantom{(1+\beta_4^{*2})^3}}
\nonumber\\&&
\frac{\mp(28c_{03}^{00}-2c^{00}_{21})\beta_4^{*3}-
(4c'_{04}-2c'_{22}+76c^{00}_{04})\beta_4^{*4}}
{(1+\beta_4^{*2})^3},
\nonumber\\
\left.\frac{\partial^2 E}{\partial\delta_4^2}\right|_{o^*}&=&
\frac{\mp15c_{03}^{00}\beta_4^{*3}-280c_{04}^{00}\beta_4^{*4}}{(1+\beta_4^{*2})^2}.
\label{e_stab4}
\end{eqnarray}
These equations are coupled, however,
since the off-diagonal derivative generally is non-zero,
\begin{equation}
\left.\frac{\partial^2 E}{\partial\beta_2\partial\delta_4}\right|_{o^*}=
\frac{2c_{12}^{1-1}\beta_4^{*2}\pm7c_{13}^{1-1}\beta_4^{*3}}{(1+\beta_4^{*2})^2}.
\label{e_stab5}
\end{equation}
Again, simplifications arise
if the coefficients of the odd powers of $\beta_4^*$ vanish, $c_{03}^{00}=c^{00}_{21}=0$.
If, in addition, the off-diagonal derivative vanishes,
$c_{12}^{1-1}=c_{13}^{1-1}=0$,
the stability in $\beta_2$ and $\delta_4$
at the hexadecapole deformation $\beta_4^*$ given by Eq.~(\ref{e_solb4})
is guaranteed by the following conditions:
\begin{equation}
(c'_{02}-c'_{20})(-c'_{02}+c'_{22})+
(2c'_{00}-c'_{20})
(2c'_{04}-c'_{22}+38c^{00}_{04})\leq0,
\label{e_cond2}
\end{equation}
and
\begin{equation}
c^{00}_{04}\leq0,
\label{e_cond3}
\end{equation}
respectively,
where use has been made of the conditions~(\ref{e_cond1}),
required to have an extremum which is stable in $\beta_4$.

Provided that $c_{03}^{00}=c^{00}_{21}=c_{12}^{1-1}=c_{13}^{1-1}=0$,
Eqs.~(\ref{e_cond1}), (\ref{e_cond2}) and~(\ref{e_cond3})
are the necessary and sufficient conditions
for the energy surface $E(\beta_2,\beta_4,\gamma_2,\gamma_4,\delta_4)$
to have a {\em local} minimum with octahedral symmetry.
It is not necessarily a unique minimum
and it may not be the {\em global} one.
In particular, the conditions do not exclude
the existence of a quadrupole-deformed minimum with $\beta_2\neq0$
at some hexadecapole deformation
which differs from $\beta_4^*$ given in Eq.~(\ref{e_solb4}).
To exclude the latter possibility for all $\beta_4^*$,
the following stronger conditions must hold
[see the first of Eqs.~(\ref{e_stab4})]:
\begin{equation}
2c'_{00}-c'_{20}\leq0,
\quad
2c'_{02}-c'_{20}-c'_{22}\leq0,
\quad
2c'_{04}-c'_{22}+38c^{00}_{04}\leq0,
\label{e_cond4}
\end{equation}
still assuming that $c_{03}^{00}=c^{00}_{21}=0$.
These stronger conditions can be combined with the inequalities~(\ref{e_cond1})
to yield, with the help of Eq.~(\ref{e_coef2a}),
\begin{equation}
c_{02}+\epsilon''_g\leq2c_{00}\leq c_{20}+\epsilon''_d,
\quad
c_{02}\leq2c_{04}+38c^{00}_{04}+\epsilon''_g\leq c_{22}+\epsilon''_d,
\label{e_cond5}
\end{equation}
where $\epsilon''_\ell$ are the scaled single-boson energies relative to the $s$ boson,
$\epsilon''_\ell\equiv\epsilon'_\ell-\epsilon'_s$.
These conditions have the advantage to be sufficiently simple
to allow an intuitive understanding.
With reference to Eqs.~(\ref{e_coef2}) and (\ref{e_coef3}),
the inequalities~(\ref{e_cond5}) express the condition that
(i) the $s$-$g$ mixing is strong enough
as compared to the energy difference $\epsilon_g-\epsilon_s$
to develop a hexadecapole deformed minimum
and (ii) the $s$-$d$ and $d$-$g$ mixing is sufficiently weak
as compared to the energy differences
$\epsilon_s-\epsilon_d$ and $\epsilon_g-\epsilon_d$, respectively,
so as the minimum to remain at $\beta_2=0$.

\begin{figure}
\centering
\includegraphics[width=5cm]{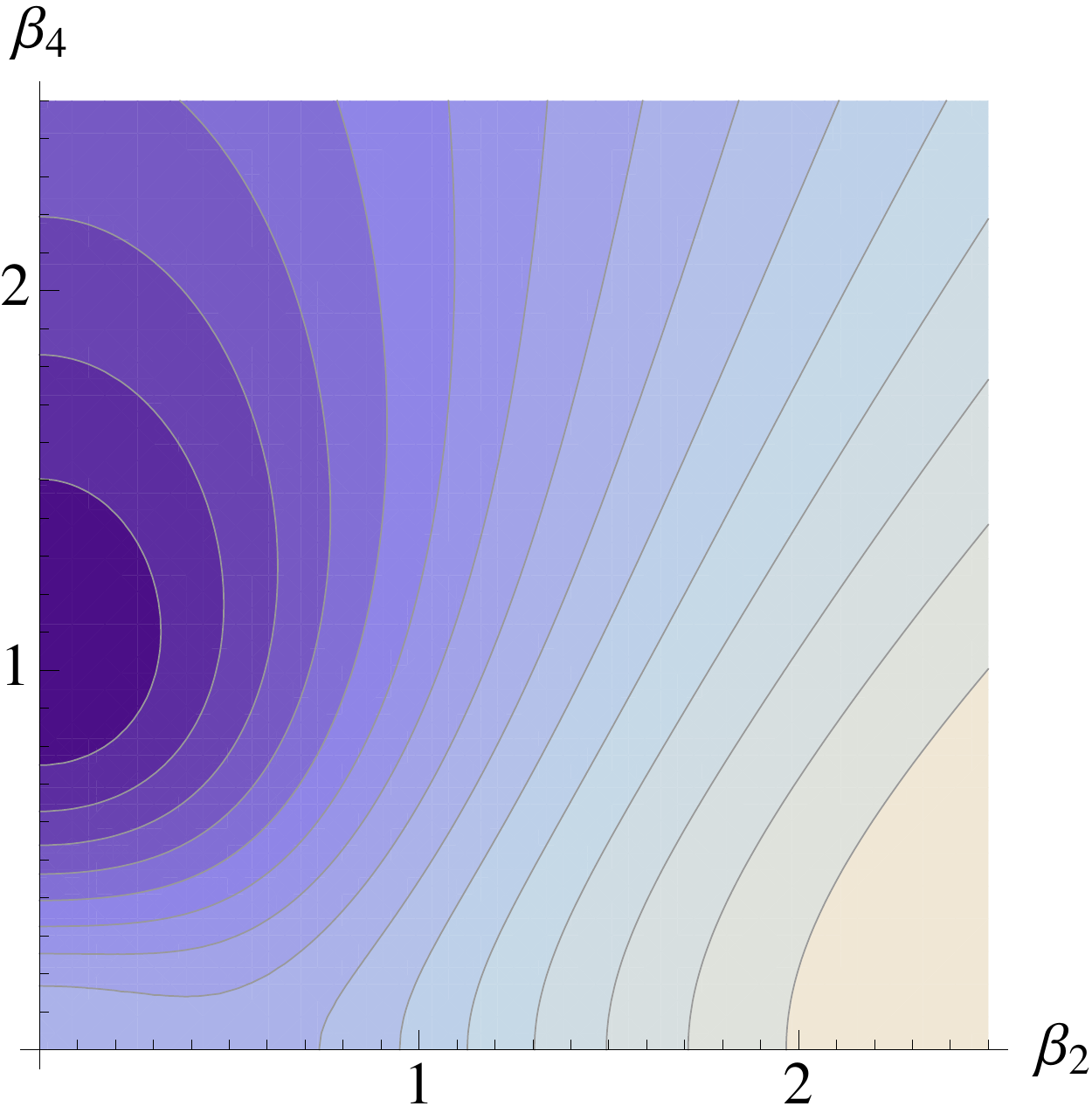}
\includegraphics[width=7cm]{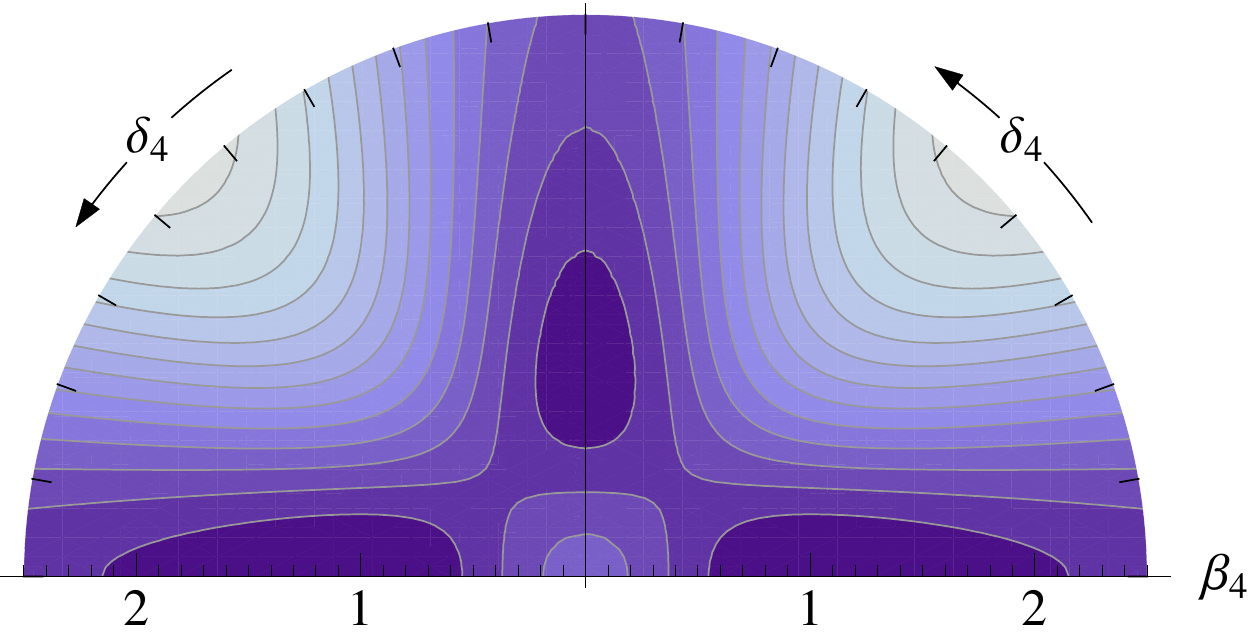}
\caption{Illustration of the energy surface
$E(\beta_2,\beta_4,\gamma_2,\gamma_4,\delta_4)$ of Eq.~(\ref{e_climit})
with coefficients
$c'_{00}=-0.2$,
$c'_{20}=-0.5$,
$c'_{02}=-1$,
$c'_{40}=0$,
$c'_{22}=-0.25$,
$c'_{04}=0.15$
and $c^{00}_{04}=-0.02$,
in arbitrary energy units times $N(N-1)$.
Left: $(\beta_2,\beta_4)$ plot for $\gamma_2=\gamma_4=\delta_4=0$.
Right: $(\beta_4,\delta_4)$ plot for $\beta_2=0$ and $\gamma_2=\gamma_4=0$.
Blue areas correspond to low energies.}
\label{f_example}
\end{figure}
In Fig.~\ref{f_example} an example is shown
of an energy surface $E(\beta_2,\beta_4,\gamma_2,\gamma_4,\delta_4)$
with coefficients in Eq.~(\ref{e_climit}) that satisfy
the conditions~(\ref{e_cond1}), (\ref{e_cond2}) and~(\ref{e_cond3}).
It is obviously not possible to display a five-dimensional surface in its full complexity
and judiciously chosen two-dimensional intersections must be shown
to illustrate the structure of $E(\beta_2,\beta_4,\gamma_2,\gamma_4,\delta_4)$.
In the study of shapes with octahedral symmetry
one may take $\gamma_2=\gamma_4=0$
since there is no dependence on these variables at the minimum.
Figure~\ref{f_example} shows two intersections,
in the $(\beta_2,\beta_4)$ plane with $\delta_4=0$
and in the $(\beta_4,\delta_4)$ plane with $\beta_2=0$,
and confirms the existence of two minima with octahedral symmetry
for $\beta_2^*=0$ and $\beta_4^*\neq0$,
with $\delta_4^*=0$ and $\delta_4^*=\pi$,
corresponding to an octahedron and a cube, respectively.
A third minimum is seen at $\delta_4^*=\pi/2$
which represents a shape with a lower discrete symmetry.

\section{Conclusions}
\label{s_conc}
In this paper a general Hamiltonian
of the \mbox{$sdg$-IBM} with up to two-body interactions
was analyzed as regards the question
of the occurrence of intrinsic shapes with octahedral symmetry.
Such an analysis requires
the use of the most general $sdg$ coherent state~(\ref{e_coh3b}),
leading to a classical energy surface of the generic form~(\ref{e_climit}).
A stability analysis of this surface leads to a set of conditions,
Eqs.~(\ref{e_cond1}), (\ref{e_cond2}) and~(\ref{e_cond3}),
which are necessary and sufficient for the occurrence of a minimum
with an intrinsic shape with octahedral symmetry.

Due to the complicated nature of the stability conditions,
only qualitative conclusions have been drawn at this point
with regard to the occurrence of octahedral shapes in the \mbox{$sdg$-IBM}.
More concrete conclusions will be drawn
in the study of a simpler \mbox{$sdg$-Hamiltonian},
which has the advantage of exhibiting dynamical symmetries
and which is the topic of another paper in this series.

\section*{Acknowledgements}
This work has been carried out
in the framework of a CNRS/DEF agreement,
project N 13760.

\end{document}